# Two-Dimensional Organic-Inorganic Room-Temperature Multiferroic

Yali Yang[1,2], Junyi Ji[1,2], Junsheng Feng[3], Shiyou Chen[2,4,*], Laurent Bellaiche[5], and Hongjun Xiang[1,2,6,*]

[1]Key Laboratory of Computational Physical Sciences (Ministry of Education), Institute of Computational Physical Sciences, and Department of Physics, Fudan University, Shanghai 200433, China

[2]Shanghai Qi Zhi Institute, Shanghai 200030, China

[3]School of Physics and Materials Engineering, Hefei Normal University, Hefei 230601, China

[4]State Key Laboratory of ASIC and System, School of Microelectronics, Fudan University, Shanghai 200433, China

[5]Physics Department and Institute for Nanoscience and Engineering, University of Arkansas, Fayetteville, Arkansas 72701, USA

[6]Collaborative Innovation Center of Advanced Microstructures, Nanjing 210093, China

**Abstract**

Organic-inorganic multiferroics are promising for the next generation of electronic devices. To date, dozens of organic-inorganic multiferroics have been reported; however, most of them show magnetic Curie temperature much lower than room temperature, which drastically hampers their application. Here, by performing first-principle calculations and building effective model Hamiltonians, we reveal a molecular orbital-mediated magnetic coupling mechanism in two-dimensional $Cr(pyz)_2$ (pyz=pyrazine), and the role that the valence state of the molecule plays in determining the magnetic coupling type between metal ions. Based on these, we demonstrate that a two-dimensional organic-inorganic room-temperature multiferroic, $Cr(h\text{-}fpyz)_2$ (h-fpyz= half-fluoropyrazine), can be rationally designed by introducing ferroelectricity in $Cr(pyz)_2$ while keeping the valence state of the molecule unchanged. Our work not only reveals the origin of magnetic coupling in 2D organic-inorganic systems, but also provides a way to design room temperature multiferroic materials rationally.

**Introduction**

Multiferroic materials, in which two or more ferroic ordering coexist, have attracted great interest due to their potential for technological applications, such as sensors, spintronics, and storage devices.[1-9] Although many inorganic multiferroics have been reported in recent years when searching for new multiferroics, most of them are not suitable for practical applications due to their weak magnetoelectric coupling, small ferroelectric polarization/magnetization or low Curie temperature.[10-12]

In particular, the organic-inorganic hybrid multiferroics, e.g., [C(NH$_2$)$_3$]M(HCOO$_3$) (M=Fe, Co, Mn, Cu, Ni), are arising as an important component of the multiferroic family in recent years,[13-18] due to their unique structural compatibility, flexibility and rich performance advantages, but they do exhibit very low magnetic Curie temperature -- which makes them difficult to meet practical demands, such as operating in room temperature.[19] It is urgent and timely to search for, or design, new room-temperature organic-inorganic hybrid multiferroics. Recently, the two-dimensional (2D) metal-organic magnets achieved a major breakthrough, *i.e.*, a layered insulating magnet [Cr(pyz)$_2$](LiCl)$_{0.7}$(THF)$_{0.25}$, where pyz = pyrazine and THF = tetrahydrofuran was synthesized by postsynthetic chemical reduction of coordination networks.[20] Hereafter [Cr(pyz)$_2$](LiCl)$_{0.7}$(THF)$_{0.25}$ will be referred to as 2·0.25(THF), as defined in ref 20. In 2·0.25(THF), each Cr bonds to four pyz rings, giving rise to a square planer coordination and neutral 2D Cr$^{II}$(pyz$^{\bullet-}$)$_2$ layers. It shows a high critical temperature (T$_c$) up to 515 K and a 7500-oersted room-temperature coercivity. Such magnetic properties made us wonder if we can design multiferroicity in this metal-organic magnet system, *i.e.*, by introducing ferroelectricity in it. Since fluorination has been shown to successfully induce or increase ferroelectricity in organic-inorganic hybrid systems,[21-25] we decided to explore if such a specific procedure can generate an electrical polarization in Cr(pyz)$_2$ system, making it a room-temperature multiferroic.

In this work, specific model Hamiltonians are proposed, in combination with first-principle calculations, to rationally design room-temperature multiferroicity in the two-dimensional organic-inorganic system Cr(pyz)$_2$. We reveal that a molecular orbital-mediated magnetic coupling mechanism determines the interactions between magnetic ions. Using such magnetic coupling mechanism, we introduce multiferroicity in Cr(pyz)$_2$ successfully by fluorinating the systems and consequently design a two-dimensional organic-inorganic room-temperature multiferroic.

**Results and discussion**

**Origin of high temperature ferromagnetism in Cr(pyz)$_2$.** To more efficiently reveal the magnetic coupling mechanism and investigate possible multiferroicity in Li$_{0.7}$[Cr(pyz)$_2$]Cl$_{0.7}$·0.25 (THF), we "exfoliate" the experimental bulk structure (with P4/mmm symmetry) to obtain a Cr(pyz)$_2$ monolayer (with P4/nbm symmetry), considering the fact that the THF molecules, as well as the LiCl layer, are neutral and nonmagnetic, therefore likely having a weak interaction with the Cy(pyz)$_2$ layer. Besides, the Cr(pyz)$_2$ monolayer structure has been shown to be stable at 400 K in a very recent work.[26] The obtained primitive cell of Cr(pyz)$_2$ monolayer is shown in Figure 1a, which contains two Cr ions and four pyz molecules. To determine the magnetic ground state of such Cr(pyz)$_2$ monolayer, we perform spin-polarized density-functional theory (DFT) calculations by setting the two Cr ions to be either ferromagnetic (FM) or antiferromagnetic (AFM). Our results show that the FM arrangement of Cr ions is lower in energy (~0.44 eV/f.u.) than that of the AFM arrangement of Cr ions. The magnetic moment of each Cr is about 3.61 μB in the FM structure. Meanwhile, each pyz molecule also shows a local magnetic moment of about 0.6 μB which aligns antiparallel with those of adjacent Cr ions, which thus leads to a ferrimagnetic ground state of Cr(pyz)$_2$ monolayer.[20] The magnetic moment of each pyz molecule originates mainly from the N atoms (~0.24 μB/N). From the structural point of view, the two pyz molecules along each diagonal direction (i.e., [110] and [1$\bar{1}$0] in Fig. 1a) are tilted in antiphase at a 47.5° angle to the [001] axis (see in Figure 1b), which is rather close to that (43.2°) of the experimental structure.[20] If viewed from the center of the primitive cell to its four corners in Figure 1a, the four pyz molecules all form a clockwise tilting pattern.

In order to gain an in-depth understanding of the magnetic coupling in the Cr(pyz)$_2$ monolayer at the orbital level, we calculate the band structure and orbital projected density of states of its ground state. As shown in Figure 1c, both the spin-up and spin-down channel of the band structure exhibit semiconducting property. The valence band maximum (VBM) and conduction band minimum (CBM) states belong to different spin channels and leads to a direct band gap of 0.59 eV. The orbital projected density of states in Figure 1d, as well as the partial charge density in Figure 1e, show that the electronic states near the Fermi level originate from the 2p orbitals of N and C atoms of the pyz molecules. The Cr 3d orbitals split into four groups [see Figure S1 in Sec. II of Supplementary Materials (SM)]: the degenerate dxz/dyz and nondegenerate dxy, dz$^2$ and dx$^2$- y² orbitals, consistent with the square planar crystal field. The occupied spin-up dxy, dxz/dyz and dz² orbitals are located at about 2.5-3 eV below the Fermi level, exhibiting in principle a magnetic moment of 4 μB/Cr. The unoccupied dx²-y² is located at 2.5 eV above the Fermi level. Thus, the FM coupling between Cr ions is mediated by the pyz molecules. We will denote it as molecular orbital-mediated magnetic coupling. Furthermore, the valence states of Cr and pyz in Cr(pyz)$_2$ monolayer are +2 and -1, respectively, same as those in the bulk structure.

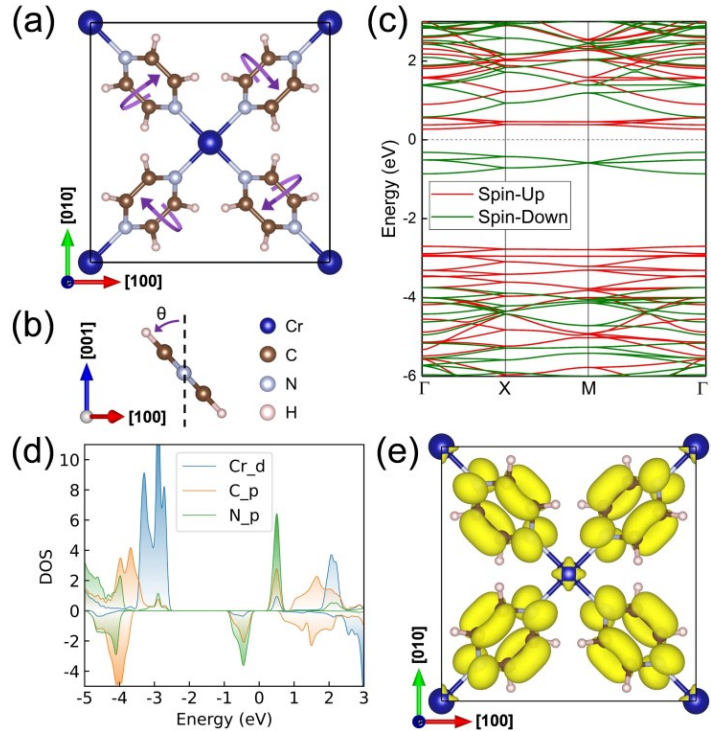

Figure 1. Structure and electronic properties of optimized Cr(pyz)$_2$ monolayer. (a) Optimized structure of Cr(pyz)$_2$ monolayer. (b) Tilted pyz molecule with respect to the [001]-axis. Purple curved arrows represent the rotations of the pyz molecule about diagonal directions. (c) Band structure of relaxed Cr(pyz)$_2$ monolayer. (d) Orbital projected density of states (DOS) of Cr(pyz)$_2$ monolayer. The Fermi level is set to zero. (e) Partial charge density of orbitals of the molecules in optimized Cr(pyz)$_2$ monolayer at an isosurface value of 0.0035 a.u. The four bands below Fermi level in (c) are used to obtain the orbitals.

To estimate the magnetic Curie temperature of Cr(pyz)$_2$ monolayer, we perform Monte Carlo (MC) simulations with a classic Heisenberg spin Hamiltonian model,

$$H = J \sum_{i<j} S_i \cdot S_j + A \sum_i S_{iz}^2,$$

where $J$ is exchange-coupling parameters between the spins and $A$ denotes the single-ion anisotropy parameters. Here, only the spins of Cr ions are considered in the effective spin Hamiltonian model. Moreover, the Cr-Cr interactions that we considered only include the nearest-neighboring Cr-Cr interactions (see the inset of Figure 2), because the interactions between the Cr ions with larger distance than the nearest-neighbors are calculated to be quite small and thus can be neglected. We adopt the four-state method[27,28] to extract the parameters $J$ and $A$, and they are calculated to be -19.77 and -0.12 meV, respectively. Here, the negative $J$ indicates the FM coupling between the Cr ions, in agreement with the recent experimental and theoretical results.[20,26] As shown in Figure 2a, the MC simulation suggests a ferromagnetic phase transition for the monolayer Cr(pyz)$_2$ at 636 K, indicating that monolayer Cr(pyz)$_2$ is a 2D room-temperature ferromagnet. This is consistent with the measured T$_c$ (515 K) of the bulk crystal[20] and higher than the value of 342 K of the monolayer structure from a recent theoretical work.[26]

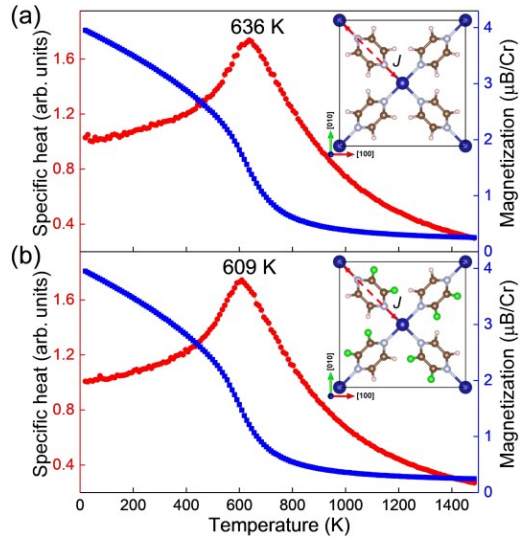

Figure 2. Specific heat (in red) and magnetic moment per Cr ions (in blue) of (a) Cr(pyz)$_2$ and (b) state-I of Cr(h-fpyz)$_2$ monolayer as a function of temperature, from Monte Carlo (MC) simulation. The inset shows the interactions between the Cr ions considered in the present spin Hamiltonian model.

In order to shed more light on the molecular orbital-mediated magnetic coupling mechanism, a triple-orbital model is constructed under the framework of the super-exchange theorem (see details in Sec. III of SM). In brief, the orbital interactions between the effective p orbitals of the organic molecule (pyz) and d orbitals of its two adjacent transition metal ions (Cr) near the Fermi level are taken into account. Our model analysis shows that, in the presence of electron hopping between orbitals, the ground state of the system changes from AFM to FM and then back to AFM as the molecular valence state changes from 0 to 2. This model not only successfully reveals that the ferromagnetic coupling mechanism between Cr ions is mediated by pyz molecules in Cr(pyz)$_2$ system, but also explains the dominant role of the valence states of the bridged molecules in the coupling between magnetic metal ions of Cr(OSO$_2$CH$_3$)$_2$(pyz)$_2$, CrCl$_2$(pyz)$_2$, Cr(pentalene)$_2$ and V(TCNE)$_2$ systems.[29-32] Thus, we can conclude that the -1 valence state of pyz molecule determines the high magnetic Curie temperature of Cr(pyz)$_2$ system. As a result, if the valence state of the bridged molecule is carefully chosen (*i.e.*, -1), the magnetic Curie temperature should be high.

**High temperature multiferroicity in fluorinated Cr(pyz)$_2$.** Let us now focus on whether multiferroicity would appear in Cr(pyz)$_2$ monolayer system. As aforementioned, fluorination can induce or increase ferroelectricity in organic-inorganic hybrid systems.[21-23] Therefore, we adopt fluorination for the Cr(pyz)$_2$ monolayer to explore the existence of multiferroicity. Usually fluorination does not change the valence state of the organic molecules, so, according to our triple-orbital model, fluorination will not change the FM ordering of Cr ions. Among the numerous possibilities of fluorination, a simple way for introducing a non-zero dipole moment is considered here: the two H atoms on the same side (*i.e.*, above or below the CrN$_4$ plane) of each pyz molecule are replaced by F atoms, while the two H atoms on the other side keeps unchanged. Consequently, fifty percent of the H atoms in Cr(pyz)$_2$ monolayer are replaced by F atoms. Hereafter, such fluorinated system will be named as Cr(half-fluoropyrazine)$_2$ [Cr(h-fpyz)$_2$]. We note that the monofluorinated substitution case (*i.e.*, only one H atom replaced by F atom of each pyz molecule) is briefly discussed in the SM.

We first construct the simplest Cr(h-fpyz)$_2$ monolayer structure (as shown in Figure 3a, named as state-I) based on the ground state of Cr(pyz)$_2$ monolayer through replacing the H atoms above the CrN$_4$ plane by F atoms. Since the F atoms have stronger electronegativity than H atoms, a dipole will be generated in each h-fpyz molecule plane with the orientation from the F atom side to the H atom side. Any of such dipoles will have both in-plane and out-of-plane components because of the non-zero tilting angle between the molecule plane and the CrN$_4$ plane. Here, the orange arrows in Figure 3 with crosses and dots indicate that the dipoles have negative and positive out-of-plane components, respectively. When summing the four dipoles in the same unit cell, one would easily find that the structure in Figure 3a shows no net in-plane dipole moment due to the antiparallel alignment of the in-plane components of the dipoles along the diagonal directions (*i.e.*, [110] and [$\bar{1}$10] direction), however, the out-of-plane dipole moment is finite and directed along the [00$\bar{1}$] direction. It means that, in such state of the Cr(h-fpyz)$_2$ monolayer, magnetism and polarization coexist, making it multiferroic.

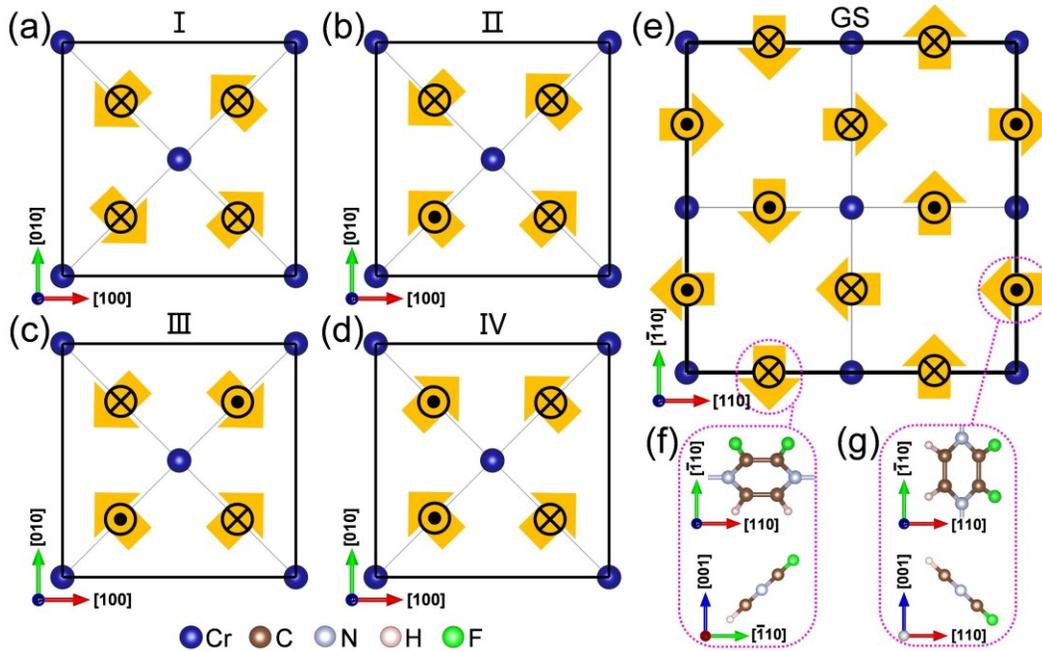

Figure 3. Illustration of different states of Cr(h-fpyz)$_2$ monolayer. (a)-(d) Four constructed states of Cr(h-fpyz)$_2$ monolayer seen from top view, named as state-I, state-II, state-III and state-IV. (e) Ground state (GS) of Cr(h-fpyz)$_2$ monolayer predicted by the MC simulation, seen from top view. (f) Side view of the h-fpyz molecules with negative and positive out-of-plane dipole components. The molecules in (f) and (g) are represented by orange arrows in (a)-(e) for clarity. The directions of the arrows indicate the in-plane directions of the dipoles of the h-fpyz molecules. The positive and negative out-of-plane components of the dipole of the h-fpyz molecules are indicated by dots and crosses within the arrows, respectively.

Considering the rotational freedom of the h-fpyz molecule, we then construct three additional Cr(h-fpyz)$_2$ monolayer structures, but with one (in Figure 3b, named as state-II) or two molecules (in Figure 3c and d, named as state-III and state-IV, respectively) in the primitive cell having a positive out-of-plane dipole component. The net dipole moment of state-II is also not zero, and has both in-plane and out-of-plane dipole moment components. In contrast, for state-III, no net dipole moment would appear. However, for state-IV, there is a net in-plane dipole moment directing along the [010] direction, with the out-of-plane component being zero. The total energy of these four relaxed states shows that state-II, III and IV is 47.66, 53.28 and 69.32 meV/f.u. lower in energy than state-I, respectively (see Table S3 in Sec. IV of the of SM).

Our above calculations therefore show that state-IV has the lowest energy among the four considered states. However, one may wonder what is the true ground state of the Cr(h-fpyz)$_2$ monolayer with the rotational degree of freedom for each h-fpyz molecule. In order to obtain such ground state in this system, we build a novel Ising-like model Hamiltonian using the Property Analysis and Simulation Package for materials (PASP)[33]:

$$H = \sum_{n=1}^{3} H_n = \sum_{n=1}^{3} \sum_{i<j} \sum_{\alpha,\beta=t,b} J_{n,\alpha\beta} \sigma_i^\alpha \sigma_j^\beta,$$

where $H_n$ indicates the $n$-th nearest-neighboring pair interaction term between molecules. Here, the nearest-, next nearest- and third nearest-neighboring interactions are considered (see in Figure 4a). Due to the large size and strong structural anisotropy of the molecule, the Ising state variables $\sigma_i^t$ and $\sigma_i^b$ in our Hamiltonian model are used to represent the H/F atoms located on the top side or the bottom side of the CrN$_4$ plane at the $i$-th molecule site, respectively. If the two H atoms on the α-side of the CrN$_4$ plane are replaced by two F atoms, $\sigma_i^\alpha$ will be assigned to -1, otherwise it will be +1. Therefore, for each two-body pair interaction, there will be four coupling parameters, i.e., $J_{n,tt}$, $J_{n,tb}$, $J_{n,bt}$ and $J_{n,bb}$ (n=1, 2, 3), as shown in Figure 4b-d. However, one would find that some of these coupling parameters are equal by symmetry, i.e., $J_{1,tt}=J_{1,bb}$, $J_{2,tt}=J_{2,bb}$, $J_{2,tb}=J_{2,bt}$ and $J_{3,tb}=J_{3,bt}$. We note that this model Hamiltonian applies only to the Cr(h-fpyz)$_2$ monolayer, a more general and complicated model Hamiltonian for the systems with different amount of fluorination is indicated in Sec. IV of SM. Different from the full complicated model in Sec. IV of SM, here only $J_{n,\alpha\beta}$ are used in the Cr(h-fpyz)$_2$ monolayer case. We note that the parameters $J_{n,\alpha\beta}$ are exactly the same as those in the complicated model, thus they are extracted together with the other parameters in the complicated model. The mapping procedure is adopted to extract the parameters, as detailed in Sec. IV of SM. We then perform Monte Carlo simulations using PASP to search for the ground state of the Cr(h-fpyz)$_2$ monolayer.

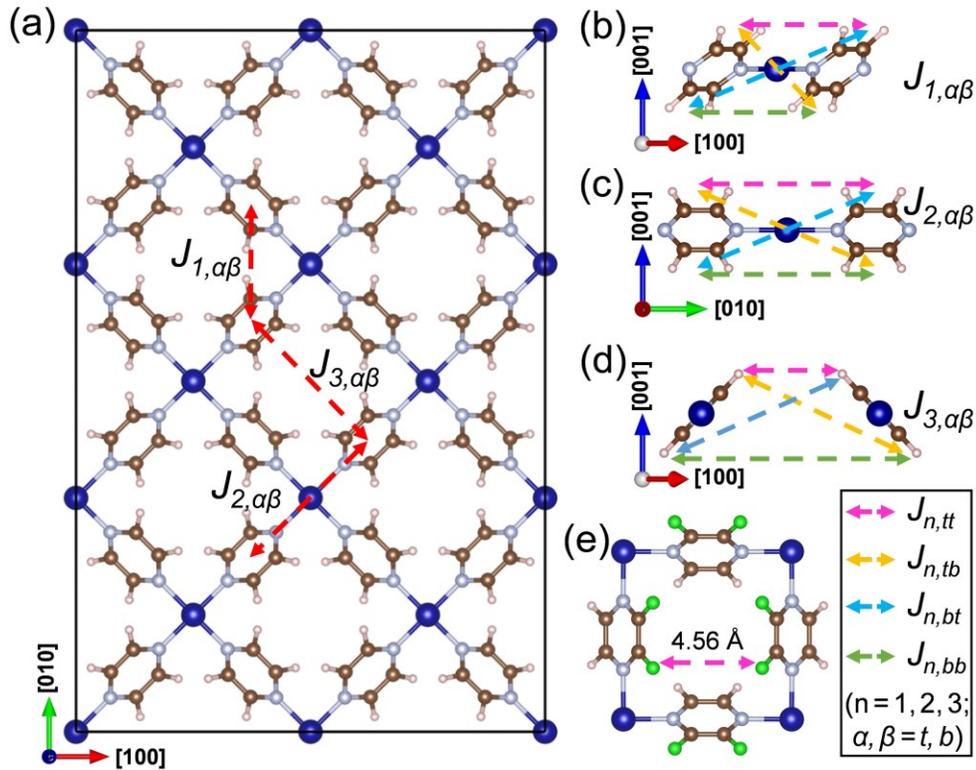

Figure 4. Schematic of the interactions between molecules considered in the MC simulation. (a) The nearest-, next nearest- and third nearest-neighboring interactions between molecules. (b)-(d) Detailed illustration of the interaction parameters $J_{n,\alpha\beta}$ between two molecules. (e) Closest distance arrangement between F atoms on the third nearest-neighboring h-fpyz molecules of optimized state-I.

The MC simulation predicts that the fluorinated molecules in Cr(h-fpyz)$_2$ arrange in the orderly manner shown in Figure 3e, making the whole monolayer forming a checkboard pattern. The overall anti-parallel alignment of the eight h-fpyz molecules in the primitive cell of the ground state (see in Figure 3e) makes the dipole moments compensated in all directions, thus no net polarization appears and it is an antiferroelectric (AFE) state. It is interesting that the ground state structure of Fig. 3e can also be regarded as made of arrays of vortices interpenetrated with arrays of antivortices. This is a special phenomenon called phase-locking.[34,35]

We then perform DFT calculation for the predicted ground state. It indeed shows that the fully optimized structure of the predicted ground state has the lowest total energy compared to the four constructed structure in Figures 3a-d with the energy difference being 83.09, 35.43, 29.81 and 13.77 meV/f.u., respectively. It is worth noting that the energies of all the five states in Figure 3 are not used for extracting model parameters $J_{n,\alpha\beta}$, but the relative energy differences between them using the Ising-like model Hamiltonian are consistent with those calculated through DFT (see Table S3 in Sec. IV of the SM) – therefore demonstrating the relevance and accuracy of our present model Hamiltonian. We note that this original Ising-like Hamiltonian, as well as our proposed magnetic coupling mechanism, is general and can be applied in principle to both two- and three-dimensional systems. Furthermore, our DFT calculations show that the ground state of $Cr(h\text{-fpyz})_2$ monolayer is a semiconductor with a direct bang gap of ~ 0.55 eV (see Sec. V of the SM).

The appearance of such checkboard pattern AFE ground state can be understood intuitively from the extracted values of the coupling parameters $J_{n,\alpha\beta}$ (see Table S2 in Sec. IV of the SM). In particular, $J_{3,tt}$ is critical in determining the AFE ground state, since it has the largest positive value, i.e., 7.33 meV, among all coupling parameters. The positive character of $J_{3,tt}$ indicates that the F atoms on the two third nearest-neighboring h-fpyz molecules prefer locating at different sides of the $CrN_4$ plane, which makes the two molecules with positive and negative out-of-plane dipole moments alternately arranged in both [110] and [$\bar{1}$10] directions (see Figure 3e). The strong coupling strength indicated by the large value (7.33 meV in magnitude) ensures the possibility of such AFE arrangement. Moreover, from the structural point of view, the length of C-F bond is larger than the C-H bond in optimized h-fpyz molecules, which would make the F atoms on the two third nearest-neighboring h-fpyz molecules close to each other if they locate at the same side of the monolayer (as shown in Figure 4e). Such close distance should increase the Pauli exclusion and electrostatic interaction energy between the $F^-$ ions, therefore, making the F atoms on the two h-fpyz molecules preferring the AFE arrangement with longer F-F distance as in the ground state of Figure 3e. Furthermore, the relatively large positive value (2.45 meV) of $J_{2,tb}$ indicate that the F atoms on the two h-fpyz molecules prefer the same side of the $CrN_4$ plane, thus making the two h-fpyz molecules having the same out-of-plane component of dipole moments. We note that there is some frustration between the pair interactions: negative $J_{1,tb}$ will prefer the configuration of state-III in Fig. 3c, while large $J_{3,tt}$ and $J_{2,tb}$ would result in the AFE ground state of the checkerboard pattern shown in Fig. 3e.

In order to estimate the magnetic Curie temperature of Cr(h-fpyz)$_2$ monolayer, we calculate the magnetic coupling strength between the nearest Cr ions as performed for the Cr(pyz)$_2$ monolayer. It is found that the fluorination of the pyz molecule has little effect on the magnetic Curie temperature, since the calculated nearest Cr-Cr magnetic coupling in the ground state of Cr(h-fpyz)$_2$ monolayer is -19.29 meV, which is very close to that of Cr(pyz)$_2$ monolayer (-19.77 meV). The coupling strength for state-I, III and IV are calculated to be -18.93, -19.18 and -19.16 meV, respectively. Those negative values indicate that the coupling between Cr ions in such fluorinated Cr(h-fpyz)$_2$ systems are all FM coupling. It, therefore, confirms the correctness of the triple-orbital model we built for revealing the magnetic coupling mechanism in Cr(pyz)$_2$ system. Furthermore, such similar magnetic coupling strengths indicate that the magnetic Curie temperature of the Cr(h-fpyz)$_2$ monolayer structures in Figure 3 should be very close to that (~636 K) of the Cr(pyz)$_2$ monolayer, far above room temperature. For example, as shown in Figure 2b, the magnetic Curie temperature of the state-I is calculated to be 609 K.

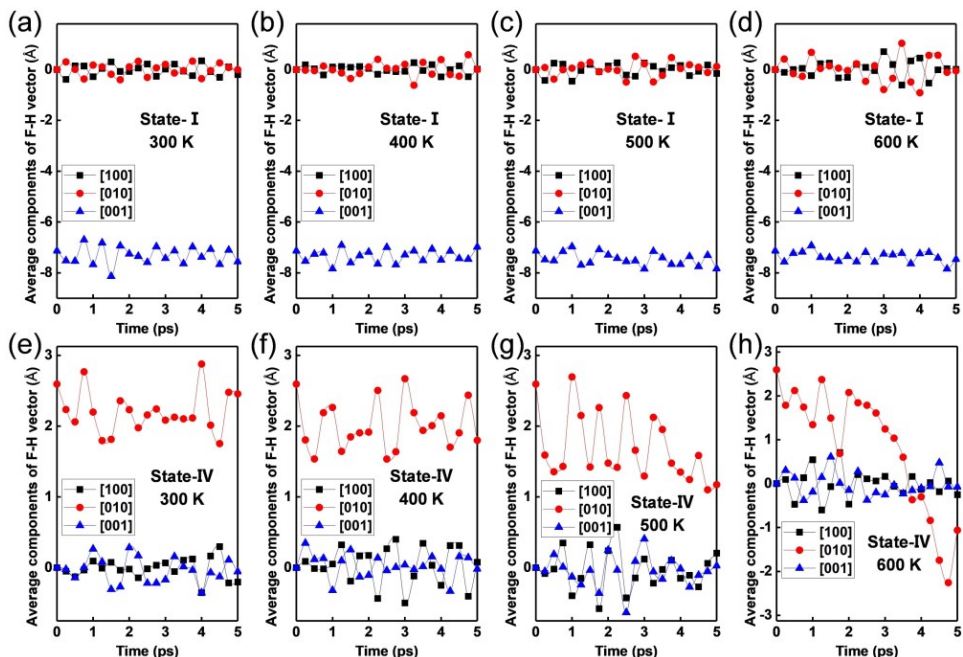

Figure 5. Thermal stability of ferroelectricity of state-I and state-IV of Cr(h-fpyz)$_2$ monolayer. (a)-(d) Fluctuation of the average components of the F-H vector as a function of simulation time for state-I of Cr(h-fpyz)₂ monolayer. (e)-(f) Fluctuation of the average components of the F-H vector as a function of simulation time for state-IV of Cr(h-fpyz)₂ monolayer.

To further check the thermal stability of the Cr(h-fpyz)₂ monolayer structures depicted in Figure 3, we perform first-principle molecular dynamics (MD) simulations (see details in Sec. VI of SM). The basic frameworks of all five Cr(h-fpyz)₂ monolayer structures of Figure 3 are found to be intact at 600 K. To further characterize the thermal stability of ferroelectricity in state-I and state-IV, we compute the average components of F-H vector, since the orientation of F-H vector is closely related to the electric polarization. Here, the F-H vector indicates the dipole vector of each h-fpyz molecule directing from the F atom side to the H atom side. The average components of the F-H vector are calculated via $\bar{d}_p = \frac{1}{N}\sum_{i=1}^{2N}[\boldsymbol{r}(F_i) - \boldsymbol{r}(H_i)] \cdot \boldsymbol{e}_p$, where the unit vector $\boldsymbol{e}_p$ represents [100], [010] or [001] direction, N is the number of h-fpyz molecules, $\boldsymbol{r}(F_i)$ [$\boldsymbol{r}(H_i)$] indicates the position of $i$-th F(H) atom. As shown in Figure 5, for state-I with the out-of-plane polarization, ferroelectricity can survive up to 600 K. While for state-IV with an in-plane ferroelectricity, ferroelectricity is stable up to 400 K, suggesting that a smaller electric field can switch the in-plane ferroelectricity at room temperature. These results show that although the ground state of Cr(h-fpyz)₂ monolayer is AFE, stable FE states can exist at room temperature. As shown in Figure S8 of SM, the electric polarization of FE state-I and state-IV of Cr(h-fpyz)₂ monolayer is evaluated to be ~4.87 μC/cm² along the [00$\bar{1}$] axis and 4.85 μC/cm² along the [010] axis, respectively. Such polarization values are smaller in magnitude than in 2D FE β-In₂Se₃ (~19.9 μC/cm²),[36] while they are comparable to the value in 2D FE BA₂PbCl₄ (~5.65 μC/cm²)[37,38] and CuInP₂S₆ (~4 μC/cm²),[39] and larger than that of $d$1T-MoTe₂ (~0.68 μC/cm²).[40] Therefore, the Cr(h-fpyz)₂ monolayer can have a large potential for the applications in multi-state memory devices. To confirm this prediction, the climbing nudged elastic band (cNEB) method[41,42] is employed to calculate the energy barrier between the ground state and state-I with the out-of-plane polarization. The calculated energy barrier is around 0.23 eV/h-fpyz (as shown in Figure 6), which is close to those of conventional ferroelectrics[43-46] and can thus be overcome under an external electric field. The energy barrier between the ground state and state-IV (having the in-plane polarization) is also calculated (see Fig. 6). The value is even smaller, ~0.15 eV/h-fpyz, which also suggests that a smaller electric field is needed to switch its ferroelectricity at room temperature.

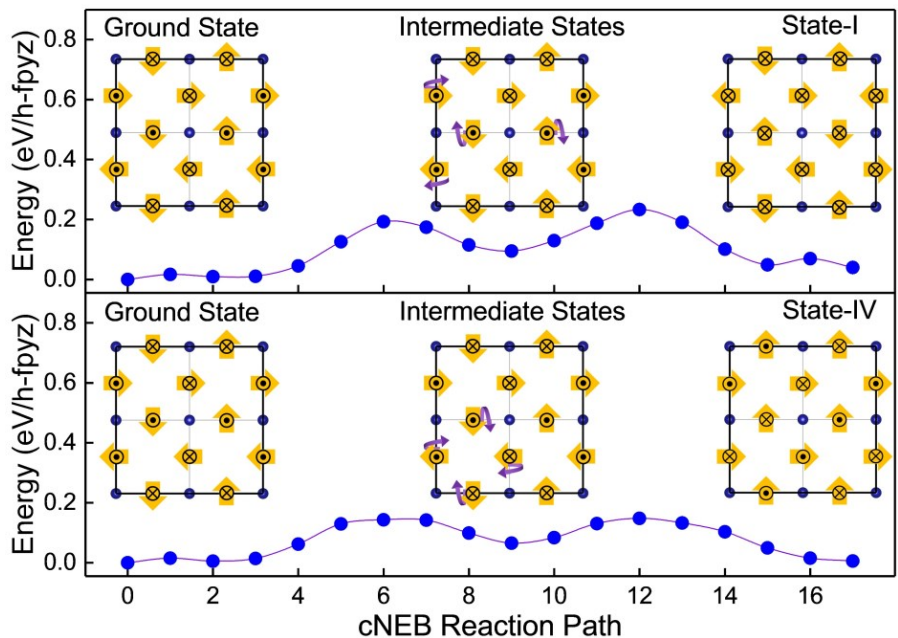

Figure 6. Transition paths between the ground state of Cr(h-fpyz)$_2$ monolayer and its state-I and IV, respectively. The h-fpyz molecules with positive and negative out-of-plane dipole components are indicated by orange arrows with dots and crosses, respectively, for clarity. The purple cycle arrows in the insets represent the rotation of h-fpyz molecules during the phase transitions.

**Conclusion**

To summarize, we achieve a rational design of a two-dimensional organic-inorganic room-temperature multiferroic. An effective triple-orbital model is built and reveals a molecular orbital-mediated FM coupling mechanism between Cr ions in the Cr(pyz)$_2$ monolayer. Using this model, the valence state of the bridged molecule between magnetic metal ions is proven to determine the coupling type between metal ions. By maintaining the valence state of the bridged molecule and taking advantage of our proposed original Ising-like model Hamiltonian, a new two-dimensional organic-inorganic room-temperature multiferroic, Cr(h-fpyz)$_2$, is designed. Its ground state is antiferroelectric, but there are stable ferroelectric states at room temperature that enable multi-state storage. This work not only sheds light on molecular orbital-mediated coupling in organic-inorganic frameworks, but also provides a guideline to design new multiferroics with high magnetic Curie temperature.

**Supporting Information**

Computational methods; Orbital projected density of states of Cr 3d orbitals in Cr(pyz)$_2$ monolayer; Triple-orbital model; Monte Carlo simulations for the Cr(h-fpyz)$_2$ monolayer; Electronic properties of Cr(h-fpyz)$_2$ monolayer; Molecular dynamics simulations for the Cr(h-fpyz)$_2$ monolayer; Electric polarization of the Cr(h-fpyz)$_2$ monolayer; Monofluorinated Cr(m-fpyz)$_2$ monolayer.

**Author Information**

**Corresponding Author**

*E-mail: chensy@fudan.edu.cn

*E-mail: hxiang@fudan.edu.cn

**Notes**

The authors declare no competing financial interest.

**Acknowledgment**

The work at Fudan is supported by NSFC (Grants No. 811825403, No. 11991061, No. 12188101, and No 12174060) and Guangdong Major Project of Basic and Applied Basic Research (Future functional materials under extreme conditions - 2021B0301030005). L.B. acknowledge the Vannevar Bush Faculty Fellowship Grant No. N00014-20-1-2834 from the Department of Defense and is thankful for support from the MonArk Quantum Foundry supported by the National Science Foundation Q-AMASE-i program under NSF award No. DMR-1906383. Yali Yang and Junyi Ji contributed equally to this work.

For Table of Contents Only

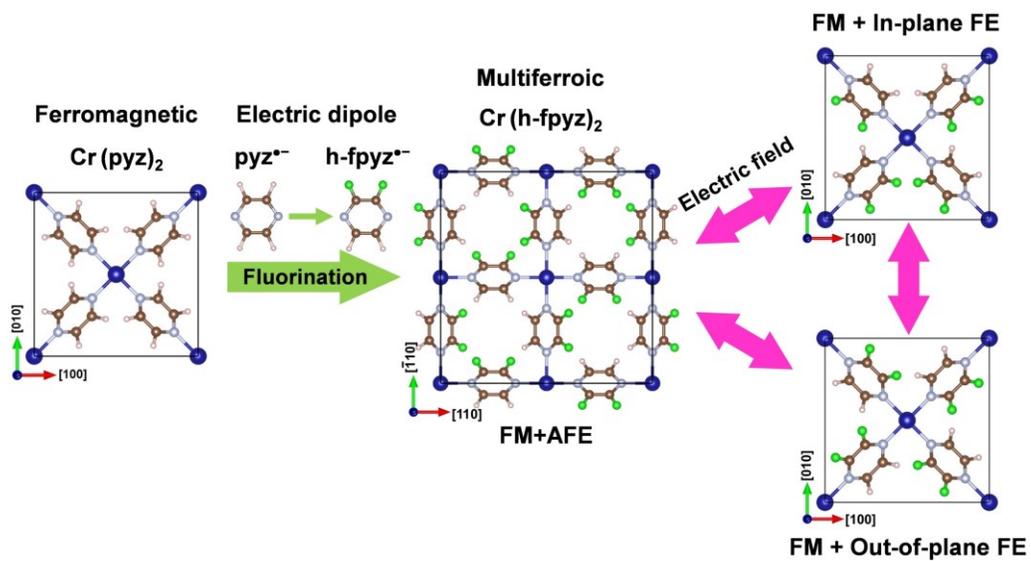